\documentstyle[12pt]{article}
\advance\voffset by 28 pt
\begin{document}
\pagestyle{empty}
\noindent
{\flushright CERN-TH.6974/93\\}
\vspace{-2mm}
{\flushright ANL-HEP-PR-93-64\\}
\vspace{2cm}
\begin{center}
{\Large \bf QCD calculation of $J/\psi + \gamma$ mass distributions}\\
\vspace{5mm}
{E.L. Berger$^{a, b)}$ and K. Sridhar$^{a *)}$}\\
\vspace{8mm}
{\em a. Theory Division, CERN, CH-1211, Geneva 23, Switzerland}\\
{\em b. Permanent Address~: High Energy Physics Division,
Argonne National Laboratory, Argonne, IL 60439, U.S.A.}\\
\end{center}

\vspace{2.cm}
\begin{abstract}
We compute the $J/\psi+\gamma$ invariant-mass distribution
from the QCD subprocess $g+g \rightarrow J/\psi + \gamma$. At large
masses, this subprocess is the dominant mechanism for $J/\psi+
\gamma$ production, and data could provide a good test of QCD.
The mass distribution peaks at relatively small masses (3.4--4.0~GeV)
and the subprocess could, therefore, represent a significant QCD background
to $J/\psi+\gamma$ decay of heavier charmonia. We also analyze
the $J/\psi$ angular distribution in the $J/\psi + \gamma$ rest frame.
\end{abstract}

\vspace{3cm}
\noindent
\vspace{1cm} $^{*)} $ sridhar@vxcern.cern.ch\\
CERN-TH.6974/93\\
August 1993\\

\vfill
\clearpage
\setcounter{page}{1}
\pagestyle{plain}
There has been much interest recently in associated production of a
$J/\psi$ and a photon ($\gamma$) in collider \cite{drekim} and fixed-target
experiments \cite{ours, mine}. In the framework of the colour-singlet
model \cite{berjon, baier}, the QCD subprocess that contributes to
the production of a $J/\psi$ and a photon is:
\begin{equation}
\label{e1}
g + g \rightarrow J/\psi + \gamma .
\end{equation}
The above subprocess is related to that for photoproduction of
$J/\psi$ by crossing. The photoproduction process and the process
in Eq.~(\ref{e1}) are gluon-initiated, and have been suggested as
sensitive probes of gluon densities in lepton-nucleon experiments
\cite{berjon}, and in hadron-nucleon experiments \cite{drekim, ours},
respectively.

The $J/\psi+\gamma$ mode is also the decay mode in which the
$P$-state charmonia are detected \cite{isr, e705}.
In this letter, we emphasise that, at low values of the
invariant mass of the $J/\psi + \gamma$ pair, the process in
Eq.~(\ref{e1}) forms an important (and computable) part of the
background to the decay of $P$-state charmonia into $J/\psi$
and $\gamma$. We present in the following the
invariant-mass distributions for the process in Eq.~(\ref{e1}).
In the mass region of the $\chi$'s, this process is a background
to the $\chi$ decay signal. At larger masses the
$J/\psi + \gamma$
continuum is dominated by the QCD subprocess and could
serve as an interesting test of QCD.

The differential cross-section for the process $g+g \rightarrow
J/\psi + \gamma$ is
\begin{equation}
\label{e2}
{d^3\sigma \over dM_{inv}dx_Tdx_1} = \cal{F} x_1G(x_1,Q^2) x_2
G(x_2,Q^2) {d\hat\sigma \over d\hat t} ,
\end{equation}
where $M_{inv}$ is the invariant mass of the $J/\psi$--$\gamma$ pair,
$x_T = 2 p_T / \sqrt{s}$ ($\sqrt{s}$ being the centre-of-mass
energy), and $G(x_{1,2})$ is the gluon distribution in the beam
and the target, respectively. The subprocess cross-section is \cite{berjon}
\begin{eqnarray}
\label{e3}
{d\hat\sigma \over d\hat t} = {16\pi\alpha\alpha_s^2 M_J \vert
R(0) \vert^2 \over 27\hat s^2} \biggl\lbrack
{\hat s^2 \over (\hat t - M_J^2)^2(\hat u - M_J^2)^2 }
 \nonumber \\
+{\hat t^2 \over (\hat s - M_J^2)^2(\hat u - M_J^2)^2 }
+{\hat u^2 \over (\hat t - M_J^2)^2(\hat s - M_J^2)^2 }\biggr\rbrack.
\end{eqnarray}
The modulus squared, $\vert R(0) \vert^2$, of the wave-function at
the origin is related to the leptonic decay width by
\begin{equation}
\label{e4}
\vert R(0) \vert^2 = {9M_J^2 \over 16\alpha^2} \Gamma(J/\psi
\rightarrow e^+e^-) = 0.544\ \mbox{\rm GeV}^3,
\end{equation}
and $\cal{F}$ in Eq.~(\ref{e2}), is given by
\begin{equation}
\label{e5}
\cal{F} = {x_1 \bar x_Tx_TM_{inv}s^2 \over 2 (x_1 - {1 \over 2}\bar x_T
e^{y_1})^3} \biggl\lbrack x_1 \bar x_T - x_1^2 e^{-y_1} - \tau e^{y_1}
\biggr\rbrack .
\end{equation}
In the above equation, $y_1$ is the rapidity of the $J/\psi$, $\tau
= M_J^2/s$, and $\bar x_T = \sqrt{x_T^2+4\tau}$. The rapidity of the
$J/\psi$ is expressed in terms of the variables $x_1$, $M_{inv}$ and
$x_T$ by the following kinematic relation~:
\begin{equation}
\label{e6}
y_1 = \mbox{\rm ln} \biggl\lbrack {(x_1x_2 + \tau) + \sqrt{(x_1x_2
+ \tau)^2 -x_1x_2 \bar x_T^2} \over x_2 \bar x_T} \biggr\rbrack ,
\end{equation}
with
\begin{equation}
\label{e7}
x_2 = {M_{inv}^2 \over x_1 s}.
\end{equation}

Using the cross-section in Eq.~(\ref{e2}), we obtain the invariant-mass
distributions by integrating over $x_1$ and $x_T$, with the
ranges of integration chosen to be
\begin{eqnarray}
\label{e8}
{M_{inv}^2 \over s} &\le x_1 &\le 1;  \nonumber \\
0 &\le x_T &\le {(x_1x_2-\tau) \over \sqrt{x_1x_2}}.
\end{eqnarray}
Note that the $x_T$ integration is not bounded from below, and
we are integrating over regions of phase space where the
photon could become soft. The subprocess cross-section in
Eq.~(\ref{e3}) is finite in the soft-photon limit.  Nevertheless,
for perturbation theory to provide a reliable answer, internal
lines in the Feynman graphs of \cite{berjon} must remain sufficiently
off-shell.  This requirement means that the kinematic region
should be chosen such that the photon does not become soft. This
condition can be ensured if we choose $M_{inv}$ such that
\begin{equation}
\label{e9}
M_{inv}^2 = M_J^2 + \Delta,
\end{equation}
where $\Delta$ is larger than some value, e.g. 1~GeV${}^2$.
If $M_{inv} \ge$~3.25~GeV, then the cross-section given in
Eq.~(\ref{e3}) can be considered a reasonable lowest-order estimate
of $J/\psi+\gamma$ production. Contributions
from soft-gluon radiation \cite{catani} will tend to increase the
cross-section, and so our estimate can be regarded as a lower
bound on the cross section arising from the gluon--gluon fusion
subprocess. A more reliable estimate of the cross-section
could be obtained after the soft-gluon corrections to
this process are computed. We plan to undertake a
study of these corrections in the future.

Integrating $x_1$ and $x_T$ over the ranges specified above,
we obtain the invariant-mass distributions for $M_{inv} \ge
3.15$~GeV. We present results for $pN$ and $\pi N$ collisions,
for typical values of $\sqrt{s}$. For the proton structure
functions, we use the MRS-S0 distributions \cite{mrs} and
for the pion distributions, the GRV-P distributions \cite{grv}.
The parton distributions are evolved to the scale $Q^2=M_J^2+p_T^2$.
In Fig.~1 we show the invariant-mass distribution $M_{inv}^3
d\sigma/dM_{inv}$ as a function of $(M_{inv}-M_J)/\sqrt{s}$, for
$pN$ collisions. The two curves correspond to beam energies
of 300~GeV and 600~GeV, respectively. We see that approximate scaling
sets in, in the region beyond $(M_{inv}-M_J)/\sqrt{s} \sim 0.07$.

In Fig.~2 we plot the number of events/GeV
for $pN$ collisions and compare with data from an
ISR experiment \cite{isr} and the E705 Fermilab experiment \cite{e705},
($\sqrt{s}=$ 62~GeV and 23.72~GeV, respectively). For the ISR data
we display the distribution as a function of $M_{inv}$, whereas
for the E705 data we use $\Delta M =
M_{inv} - M_J$. In comparing the shapes of our curves with the data, we
normalise to the experimental points at $M_{inv}=$~4.8~GeV and $\Delta
M=0.88$~GeV, respectively. The shape of the distributions away from
the resonance region may be in reasonable agreement with the
data, but it is desirable to have data that extend to larger values
of $M_{inv}$. This experimental information will be valuable since
it will serve as a test of the QCD predictions of both the
magnitude and the $M_{inv}$ dependence.  We suggest that data on
associated production of $J/\psi + \gamma$ at large invariant masses
could provide a QCD test of gluon initiated hard-scattering reactions
similar in quality to the QCD test that massive lepton-pair production
provides for quark-antiquark initiated hard-scattering.

With the normalisation fixed as stated in the paragraph above, we find that
the QCD process could
account for about 25\% of the events under the resonance, for both
sets of data. In \cite{isr,e705} the background to $\chi$ decay is
constructed from uncorrelated $J/\psi$ and $\gamma$ momentum vectors
obtained from different events and normalized to the data above the
$\chi$ mass region.  Thus, in the analyses of \cite{isr,e705}, the QCD
signal that we are discussing seems to have been subsumed into the
"background".

In Fig.~3 we show the invariant-mass
distributions for both $pN$ and $\pi N$ collisions at $\sqrt{s}=$
23.72~GeV.  A lower cut on the photon energy is often employed in the data
because definition of the photon may be poor at low photon energies.
We present results for three different cuts on the
laboratory energy of the photon ($E_{\gamma}^{\mbox{\rm lab}} \ge$ 0,\
1.0,\ 2.5~GeV). The mass distribution is not very sensitive to the
photon energy cut, although the peak is shifted slightly to larger
$M_{inv}$ as the photon energy cut is raised.

In addition to the mass distribution discussed above, we have
analysed the angular distributions of the $J/\psi$ in the
$J/\psi$--$\gamma$ rest-frame (which corresponds, in lowest order
perturbation theory, to the subprocess centre-of-mass frame). The
angular distribution is expressed as
\begin{equation}
\label{e10}
{d^3\sigma \over d\mbox{\rm cos}\theta^* dM_{inv}dx_1} =
{(\hat s- M_J^2) \over x_1 M_{inv}} x_1G(x_1,Q^2) x_2
G(x_2,Q^2) {d\hat\sigma \over d\hat t},
\end{equation}
with $p_T$ given in terms of the polar angle $\theta^*$ of the $J/\psi$ as
\begin{equation}
\label{e11}
p_T = {(\hat s- M_J^2) \over 2 M_{inv}} \mbox{\rm sin} \theta^*.
\end{equation}
After integrating over $x_1$, we obtain $d\sigma/d\mbox{\rm cos}
\theta^* dM_{inv}$. In Fig.~4 we show the angular distribution as a function
of $\mbox{\rm cos} \theta^*$ for different values of $M_{inv}$, for
$pp$ collisions at $\sqrt{s}=23.72$~GeV. The distribution is
symmetric about $\mbox{\rm cos}\theta^* = 0$, and the
dependence on $\mbox{\rm cos}\theta^*$ becomes stonger with
increasing $M_{inv}$.

Experimental information on angular distributions in the $J/\psi$--$
\gamma$ rest system should be useful in establishing the production
mechanism. In the decay of $\chi_2$'s to $J/\psi + \gamma$, the
$J/\psi$ angular distribution is expected to follow \cite{ioffe}
a $(1+\mbox{\rm cos}^2 \theta^*)$ behaviour. In the same mass
region ($M_{inv}=3.5$~GeV), the QCD angular distribution that we
have computed obeys a $(1+\alpha\mbox{\rm cos}^2 \theta^*)$ fit,
with $\alpha=0.48$.  Selections on $\mbox{\rm cos}\theta^*$
should help to enhance the resonance signal relative to the QCD process
and vice-versa.   For larger $M_{inv}$, the QCD angular distribution
shows a large contribution from an additional $\mbox{\rm cos}^4 \theta^*$
term.  In the $\chi$ resonance region, the resonance and part of the QCD
background occur in the same partial wave.  A complete analysis of this
mass region would require a proper treatment of the effects of
final-state interactions.\cite{basber}

In summary, we have presented calculations of invariant-mass distributions
for associated production of $J/\psi+\gamma$ through gluon--gluon fusion. This
QCD process provides a potentially important background to the
$J/\psi+\gamma$ decay mode of the $P$-state charmonia. Angular distributions
may be useful for separating the QCD contribution from the charmonium
resonance signal.  At large invariant masses, associated production of
$J/\psi+\gamma$ should serve as a good test of hard-scattering initiated by
gluons.

\newpage
\section*{Figure captions}
\renewcommand{\labelenumi}{Fig. \arabic{enumi}}
\begin{enumerate}
\item   
The scaling distribution
$M_{inv}^3 d\sigma/dM_{inv}$ as a function of $(M_{inv}-M_J)/\sqrt{s}$,
for $pp$ collisions, with beam energy = 300~GeV (solid curve) and
600~GeV (dashed curve).

\item   
Comparison of the mass distributions with data from ISR \cite{isr}
(upper figure) and E705 \cite{e705} (lower figure). The theoretical
curves are normalised to the data sets at $M_{inv}=4.8$~GeV in the
upper figure, and at $\Delta M=0.88$~GeV in the lower figure.

\item   
The mass distribution $d\sigma/dM_{inv}$ at $\sqrt{s}=23.72$~GeV,
as a function of $M_{inv}$, for $pp$
collisions (left figure), and $\pi p$ collisions (right figure).
The solid, dashed and dash-dotted curves correspond to cuts
on the laboratory energy of the photon ($E_{\gamma}^{\mbox{\rm
lab}} \ge$ 0,\ 1.0,\ 2.5~GeV).

\item   
The angular distribution
$d\sigma/d\mbox{\rm cos} \theta^* dM_{inv}$ as a function of
$\mbox{\rm cos} \theta^*$. The curves (from top to bottom) correspond
to $M_{inv}=$ 3.5, 4.5, 5.5, 7.5 and 9.5 GeV.

\end{enumerate}
\end{document}